\documentclass[twocolumn, %reprint,
superscriptaddress,
 amsmath,amssymb, prl,
 aps,
]{revtex4-1}

\usepackage{graphicx}% Include figure files
\usepackage{dcolumn}% Align table columns on decimal point
\usepackage{bm}% bold math
\usepackage{color}

\begin{document}

\title{Effect of quantized conductivity on the anomalous photon emission radiated from atomic-size point contacts}

\author{Micka\"el Buret}
\affiliation{Laboratoire Interdisciplinaire Carnot de Bourgogne CNRS UMR 6303, Universit\'e de Bourgogne Franche-Comt\'e, 21000 Dijon, France
}%

\author{Igor V. Smetanin}%
\author{Alexander V. Uskov}
\affiliation{Lebedev Physical Institute, Leninsky pr. 53, 119991 Moscow, Russia
}
\author{G\'erard Colas des Francs}
\affiliation{Laboratoire Interdisciplinaire Carnot de Bourgogne CNRS UMR 6303, Universit\'e de Bourgogne Franche-Comt\'e, 21000 Dijon, France
}%
\author{Alexandre Bouhelier}
\email{alexandre.bouhelier@u-bourgogne.fr}
\affiliation{Laboratoire Interdisciplinaire Carnot de Bourgogne CNRS UMR 6303, Universit\'e de Bourgogne Franche-Comt\'e, 21000 Dijon, France
}%

\date{\today}% It is always \today, today,

\begin{abstract}
We observe anomalous visible to near-infrared electromagnetic radiation emitted from electrically driven atomic-size point contacts. We show that the number of photons released strongly depends on the quantized conductance steps of the contact. Counter-intuitively, the light intensity features an exponential decay dependence with the injected electrical power. We propose an analytical model for the light emission considering an out-of-equilibrium electron distribution. We treat photon emission as bremsstrahlung process resulting from hot electrons colliding with the metal boundary and find a qualitative accord with the experimental data.

\end{abstract}

\maketitle

%\tableofcontents

An atomic-scale contact formed between two macroscopic electrodes has been a canonical testbed for understanding the quantum nature of electron and heat transport at this ultimate length scale~\cite{Ruitenbeek03,Reddy17}. Central to the discussion is the role of dissipation, which must be taken into account in any finite conductance externally driven electrical device. In the phenomenological treatment of quantum transport of a one-dimensional conductor~\cite{datta_1995,Beenaker96}, the collision-free transmission imposes the dissipation to occur away from the ballistic channel, \textit{i.e.}, in the reservoirs contacting the conductor in a distance equals to the inelastic electron mean free path. Even when describing electron flow from first-principle quantum kinetics~\cite{Das05}, inelastic coupling to the interface region guarantees the conservation of the charge required for any open geometry~\cite{Green09}. It is generally understood that the main channel for energy dissipation in a out-of-equilibrium ballistic contact occurs via a coupling to the phonon bath and the local generation of heat~\cite{Todorov98}. Population of the phonon distribution has been confirmed in voltage-dependent conductance spectroscopies~\cite{Vieira02,Jauho04} and in weak-field current fluctuations analysis~\cite{Reznikov95,Ruitenbeek12}. Such inherent fluctuations of the charge current is necessarily accompanied by the emission of transverse electromagnetic field. For low driving voltages, \textit{i.e.}, in the linear regime, the radio-frequency photons may feature nonclassical statistics depending on the voltage applied~\cite{Schomerus04,Blatter10} and the temperature~\cite{Beenaker10}. This has been experimentally measured on tunnel junctions at cryogenic temperature and emitting in the GHz frequency range~\cite{Portier10,Reulet16}. For larger driving bias, the situation complicates and the standard fluctuation-dissipation theory is no longer applicable~\cite{Rogovin74}. Electron-electron scattering must be included in the dissipation as it contributes to elevate the temperature of the Fermi-Dirac distribution. In turns, the electron and the phonon subsystems are not longer thermalized~\cite{Federovich00,Green04,Diventra06,NatelsonSR14}. Here, we identify the presence of a corollary dissipation mechanism. We show that the high-temperature nonequilibrium electron gas formed in an externally-driven atomic-scale contact is dissipating energy by emitting electromagnetic radiation tailing in the visible part of the spectrum. We observe an increase of the photon rate every time a transmission channel governing the electronic transport closes. Opposite to the conventional exchange of energy to a thermal bath and to standard electroluminescence, the light intensity emitted by the contact inversely scales with the electrical power dissipated nearby the ballistic conductor. We treat photon emission as spontaneous bremsstrahlung radiation emerging when hot electrons collide with the metal wall to explain the experimental results.

In this work, atomic-size electron channels are formed by electromigrating Au constrictions~\cite{mceuen99}. Quantized steps of the conductance in units of the quantum of conductance $G_0=2e^2/h$ is the signature of a ballistic transport, where $e$ is the electron charge and $h$ is Planck's constant~\cite{Strachan2005,HoffmannAPL08}. We detect the light activity during the electromigration process by capturing photon emission with two cross-polarized single photon counting avalanche photodiodes (APD). The quantum efficiency of the APDs sets the detected spectral range to high energy photons spanning the visible and near-infrared region (ca. 1.2~eV-3.1~eV). The detail of the experimental methodology is provided in Supplemental Material~\cite{supp}. Figure~\ref{fig:apdvst} shows two examples of time traces recorded towards the end of the electromigration process leading to the electrical failure of the devices. The applied bias is constant at $V_{\rm b}=0.8$~V in (a) and $V_{\rm b}=0.7$~V in (b). The step-like evolution of the normalized conductance $G/G_0$ suggests that the devices undergo change of transport mechanism from ballistic to tunnel; the abrupt passage is taking place at $t=27.8$~s in Fig.~\ref{fig:apdvst}(a) and at $t=64.2$~s in Fig.~\ref{fig:apdvst}(b). The relative large values of $V_{\rm b}$ reduced the probability of $G$ to explore the smallest integer numbers $N\times G_0$~\cite{Sakai04}, and the last measured steps are at $N=4$ in both cases. Figure~\ref{fig:apdvst} also displays the simultaneously acquired photon counts measured by the APDs. The graphs show an unambiguous correlation between the conductance steps and the light emission. Photons emitted in the detected spectral window are measured as soon as $G \sim 5G_0$ with a constant rate during the conductance plateaus. A ten fold increase of the number of photons is concomitant to the closing of an electron transmission channel identified by the short excursion of $G$ at $4G_0$ in both examples. Immediately after the rupture of the device, the tunneling junctions in Fig.~\ref{fig:apdvst}(a) and Fig.~\ref{fig:apdvst}(b) have conductances of $G=6\times 10^{-3}G_0$ and $G=1.4\times 10^{-3}G_0$, respectively. In both cases, the photon rate drops when transport changes from ballistic to tunnel. During the entire time traces and the excursion of $G$ in the different transport regimes, the photon energy is always greater than the bias energy. The quantum inequality $h\nu  \leq eV_{\rm b}$ is systematically violated, where $\nu$ is the frequency of the photon. We can thus exclude emission processes akin to inelastic tunneling~\cite{Novotny-17} to  explain the light activity. This is further confirmed by the similarity of the signals detected by the two cross-polarized APDs. Inelastic coupling to raditaive surface plasmon modes are expected to show a polarization anisotropy~\cite{Parzefall15,Uskov_ACS17}.

 \begin{figure}
\includegraphics[width=1\columnwidth]{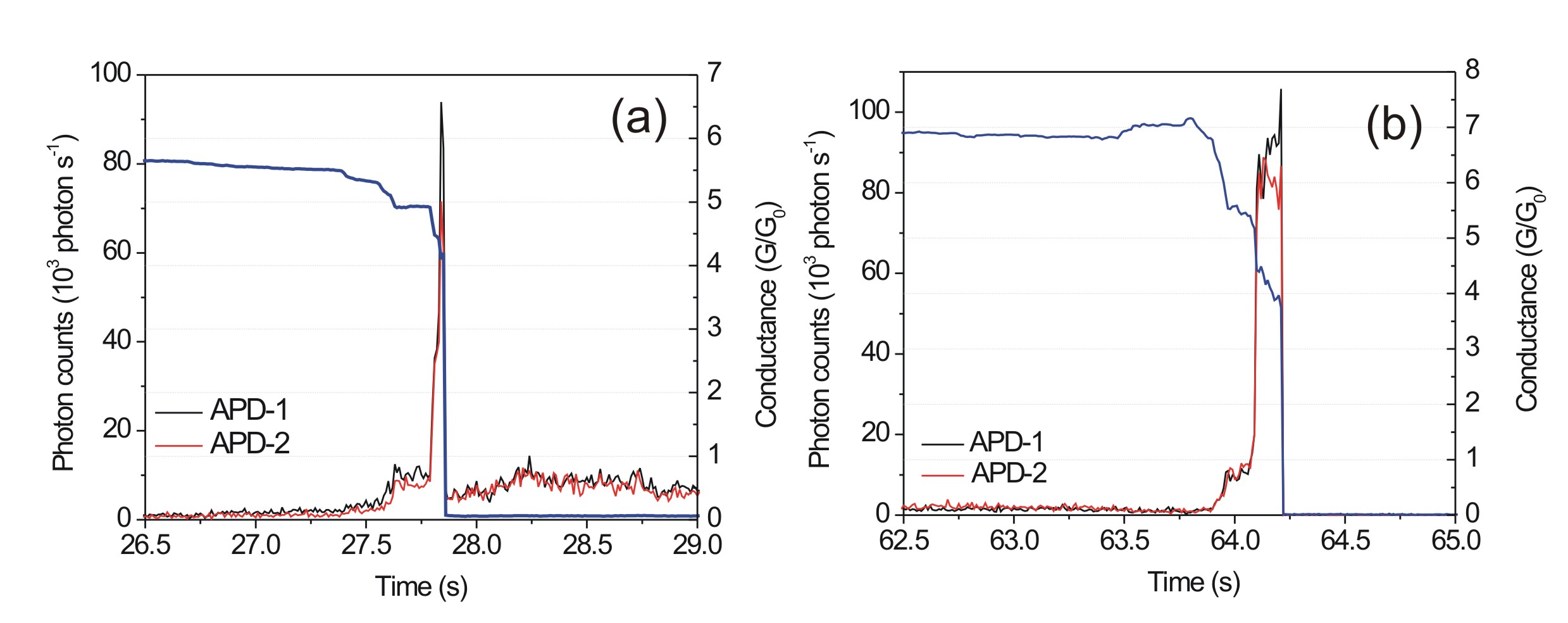}
\caption{\label{fig:apdvst} (a) and (b): Time traces showing the conductance and photon counts captured during the last moment of electromigration for two different devices. The rupture occurs at $t=27.8$~s in (a) and at $t=64.2$~s in (b). The conductance is normalized by the quantum of conductance $G_0$.}
 \end{figure}

 \begin{figure}
\includegraphics[width=1\columnwidth]{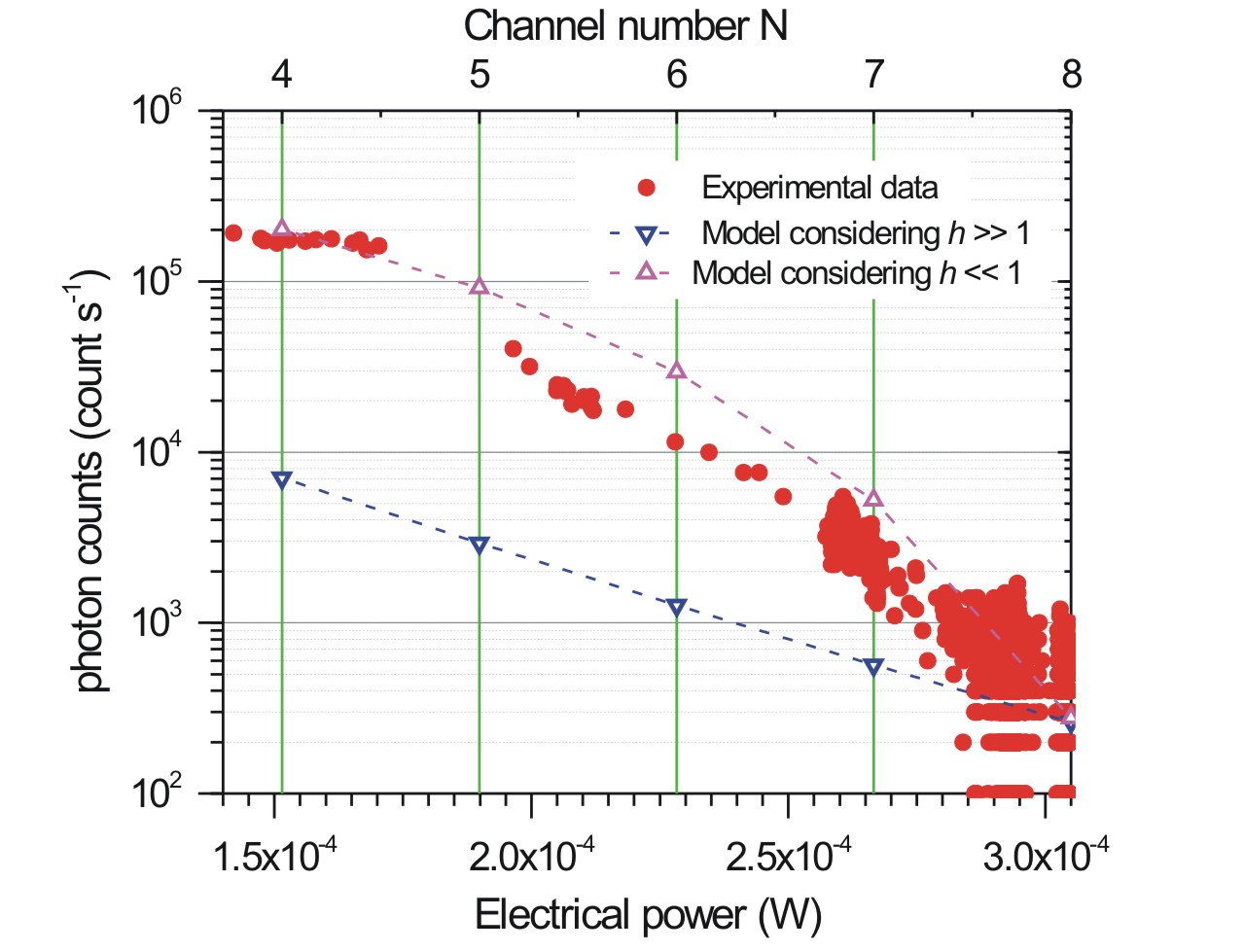}
\caption{\label{fig:apdvsG} Semi-logarithmic plot of the photon count dependence on the dissipated power $P=I_{\rm b}^2\times G^{-1}$ in the contact ($V_{\rm b}=$700~mV). The magenta and the blue inverted triangles are the model expectation considering either a vanishing heat exchange at the side wall ($h<<1$) or an efficient thermalization ($h>>1$), respectively.}
 \end{figure}

Early observations of an overbias emission in an atomic contact has been shown to follow a power law relationship with the electrical power injected in the device. For a given value of the conductance, and regardless of the emission mechanism at play, increasing the current by changing the electrical bias drastically boosted the detected photon counts~\cite{Schull09,Dumas16}. In the present experiment, the voltage bias is maintained at a constant value during the last moment of electromigration. The excursion of the conductance in the ballistic regime allows us to monitor the evolution of the photon counts with the electrical power dissipated in the contact without changing the driving conditions, and to obtain a deeper insight on the emission mechanism. When transport channels are closing, the electrical power dissipated in the contact reduces concomitantly. Figure~\ref{fig:apdvsG} shows a semi-logarithmic plot of the measured light intensity (red circles) versus the electrical power $P$ inferred from Fig.~\ref{fig:apdvst}(a) before the electrical failure using the relation $P=I_{\rm b}^2\times G^{-1}$, where $I_{\rm b}$ is the current flowing through the contact. The graphs unequivocally demonstrates that the photon counts is maximum at lower electrical power and features an exponential decay with $P$ before saturating aroudn the lowest electrical power. This trend is opposite to measurements performed at constant $G$~\cite{Schull09,Dumas16}. This trend has been consistently confirmed with other devices (see Supplemental Material~\cite{supp})

In the following, we develop a theoretical framework to understand the relationship between the number of channels opened for electron transmission and the optical activity emitted at an overbiased photon energy. The delivered  electric power scales with the number of transport channels $N$ as $P_{N}=V_{\rm b}^2 N G_0 $. The radius of the $N$-th channel can be estimated as $r_{N}\approx N r_1$, where $r_1\approx \lambda_{F}/4$ is the characteristic radius of the first quantum channel~\cite{datta_1995} and $\lambda_{F}$ is the Fermi wavelength of the ballistic electrons. As a result, both current and power densities are increasing in proportion of $1/N$ when channels are closing and results in a rise of the peak electron temperature within an area located at the end of the transport channel. Photons may be emitted by such a nonequilibrium distribution if electrons interact with the surface~\cite{Federovich00}. Qualitatively, this is expected to be the origin of the measured increasing photon yield at energies higher than the bias when the constriction explores the lower values of conductance quanta.

Below, we present a simple qualitative model which illustrates the above consideration. We assume that the electric current is transported by a channel connected to the drain contact through an interconnection region which we model by a cylinder of radius $R_0$ and finite length $L_{c}$ along the $z$ axis. The electron subsystem in this interconnection region is out of equilibrium due to the fast heating with arriving and colliding quasi-ballistic electrons. We assume a local electron temperature $T_e$, which is well above the homogeneous lattice temperature $T_{L}$~\cite{NatelsonSR14}. We treat the heat transport problem in this interconnection region in the frame of the two-temperature model, assuming the lattice temperature $T_L$ does not change significantly along $L_{c}$.

In accordance with the experimental conditions, we seek a steady state temperature distribution

\begin{equation}
C_e\frac{\partial T_e}{\partial t}=\nabla (\kappa_e \nabla T_e)-g(T_e-T_{L})=0,
\end{equation}
where $g$ is the electron-lattice coupling constant, $\kappa_e$ and $C_e$ are the electron thermal conductivity and heat capacity. As the natural boundary conditions, we assume the electron temperature in the drain electrode far from the contact to be at the equilibrium with the lattice temperature, so that
$T_e (z=L_{c})=T_{L}$. At the front end of the contact $z=0$, the electric power is assumed to be homogeneously deposited in a spot with the radius $r_{N}$ of  opened quantum transport channels $N$, so that the boundary heat flux is $\kappa_e(\partial T_e/\partial z)_{z=0}=-p_{N}\Theta(r-r_{N})$, where $p_N=P_N/\pi r_N^2=(G_0 V_{\rm b}^2/\pi r_1^2) N^{-1}$, and $\Theta(u)$
is the step function defined as $\Theta (u>0)=0$ and $\Theta (u<0)=1$. At the side wall of the cylinder $r=R_0$,  we assume that the heat flux is determined by the energy loss of the electrons in collisions with the metal boundary, in analogy with the Fedorovich-Tomchuk mechanism~\cite{tomchuk90} (See Supplemental Material~\cite{supp} for the discussion). In this framework, the heat flux at the side wall is proportional to the squared temperature $\kappa_e (\partial T_e / \partial r) = - B(T_e^2-T_L^2)$.
Here B is a proportionality coefficient. As far as $\kappa_e= b \times T_e$, we find
$(b/2) (\partial T_e^2 / \partial r)= - B(T_e^2-T_L^2)$. Finally, we set $h=2BR_0/b$ and find
$\partial T_e^2/\partial r+(T_e^2-T_L^2)h/R_0=0$ with $h$ is the dimensionless parameter characterizing the electron energy exchange rate at the side wall. The limit $h=0$ corresponds to zero heat flux at the side wall and $\partial T_e^2 / \partial r$ vanishes. Large values of the parameter $h>>1$ describe a fast energy exchange leading to rapid establishment of the equilibrium between electrons and the lattice, \textit{i.e.} $T_e=T_{\rm L}$.

Under the above assumptions, the steady-state temperature distribution is written as
\begin{equation}
T_e^2=T_L^2+T_0^2 \sum_{n=0}^{+\infty} a_nJ_0(\mu_n r/R_0) \sinh (\lambda_n(L_c-z)/R_0)
\label{Eq:2}
\end{equation}

The coefficients $a_n$ are
\begin{eqnarray}
a_n & = &\frac{\mu_nJ_1(\mu_n \zeta_N)}{\lambda_n\mu_n^2([J_1(\mu_n)]^2+[J_0(\mu_n)]^2)} \frac{1}{\cosh(\lambda_nL_c/R_0)} \nonumber \\
& = &\frac{\mu_nJ_1(\mu_n \zeta_N)}{\lambda_n(\mu_n^2+h^2)[J_0(\mu_n)]^2}\frac{1}{\cosh(\lambda_nL_c/R_0)}
\end{eqnarray}

Here $\zeta_N=r_N/R_0$, $\mu_n$ is the root of the equation $\mu_nJ_1(\mu_n)/J_0(\mu_n)=h$ with $n=0, 1, 2$,~\ldots and the eigen values of the problem along the $z$ axis are $\lambda_n=(\mu_n^2+R_0^2/L_0^2)^{1/2}$. $L_0$ is a characteristic length defined in the Supplemental Material~\cite{supp}. The coefficient $T_0^2$ before the sign of the sum in Eq.~\ref{Eq:2} doesn't depend on the channel's number, $k_B^2T_0^2=96(eV_{\rm b})^2 E_F/\pi^4\hbar N_e v_F^2\lambda_F \tau$. Within the accepted above approximation for the electron transport relaxation time, $\tau \sim R_0/v_F$, using the well-known relation $N_e=k_F^3 /3\pi^2$, we find $k_BT_0=(6/\pi^2)(eV_{\rm b})(\lambda_F/R_0)^{1/2}$, which for the applied voltage $V_{\rm b}=0.7$~V results in $T_0 [{\rm 10^3~K}]\approx 4.92\sqrt{\lambda_F/R_0}$.

According to Eq.~\ref{Eq:2}, the maximum temperature is in the center of a hot spot situated at the front end of the cylinder ($ z=0$, $r=0$). When the side wall heat transfer is fast ($h\rightarrow +\infty$), the boundary condition reads  $T_e(r=R_0)=T_L$, and $\mu_n \rightarrow \mu'_n$ becomes zeros of the Bessel function $J_0(\mu'_n)=0$ with $n=1,2~\ldots$. Because $L_0/R_0\gg 1$~\cite{supp}, $\lambda_n \approx \mu'_n$ and the maximum temperature in this limiting case is
\begin{equation}
T_{e [h\rightarrow +\infty]}^2 \approx T_L^2+T_0^2 \sum_{n=0}^{+\infty} \frac{J_1(\mu'_n \zeta_N)}{[\mu'_nJ_1(\mu'_n)]^2} \tanh \left(\frac{\mu'_nL_c}{R_0}\right)
\end{equation}
In the opposite case of vanishing energy exchange at the side wall,  $h<<1$ or even $h=0$, the roots $\mu_n \rightarrow \mu''_n$ consist in a zeroth root $\mu''_0\approx \sqrt{2h}$ (for which the corresponding eigen value is $\lambda_0 \approx (2h+R_0^2/L_0^2)^{1/2} \rightarrow R_0/L_0$), and the sequence of the roots of the first-order Bessel function $J_1(\mu''_n)\approx 0$, with $n=1, 2\ldots$, and the maximum temperature can be estimated as
\begin{equation}
T_{e[h\rightarrow 0]}^2 \approx T_L^2+T_0^2 \left[ \frac{\zeta_N L_c}{R_0}+\sum_{n=0}^{+\infty} \frac{J_1(\mu''_n \zeta_N)}{[\mu''_nJ_0(\mu''_n)]^2} \tanh \left(\frac{\mu''_nL_c}{R_0}\right)\right]
\label{h0}
\end{equation}

\begin{figure}[h]
\includegraphics[width=1\columnwidth]{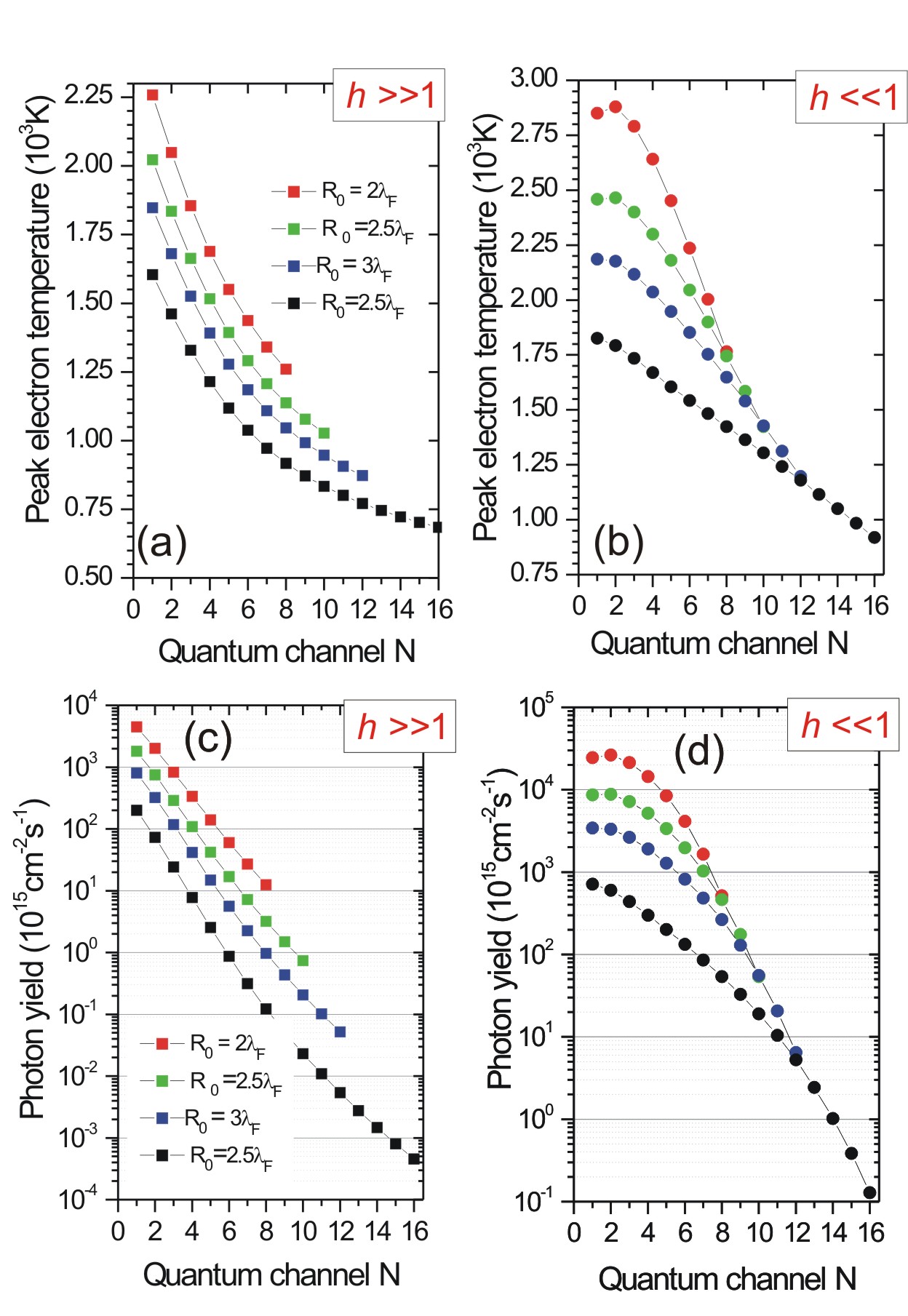}
\caption{\label{fig:peakTvsN1} (a) and (b) Peak electron temperature versus the number of open quantum channels $N$ in the contact calculated at fixed cylinder length $L_c=\lambda_F/2$. (c) and (d) are the corresponding thermal bremsstrahlung radiation rates. Square points represent data at large parameter $h>>1$, while circle points correspond to $h<<1$. Data are plotted for different radii of the cylinder interconnection region. 
}
 \end{figure}

Figure~\ref{fig:peakTvsN1}(a) and (b) displays the dependence of $T_e$ versus the number of quantum transport channels at the fixed length $L_c =R_0/3$ and various radii $R_0$ ranging from $R_0 =2\lambda_F$ (8 quantum channels) to $R_0 =4\lambda_F$ (16 quantum channels available) and for the two heat exchange scenarios. Clearly, $T_{e}$ drops when more channels are available for electron conduction. For each channel number $N$, the peak temperature is greater with a decrease of the interconnection radius. The dependencies are more pronounced when $h << 1$ and for smaller radii and become smoother with an increase in $R_0$. The dependence of $T_e$ with fixed $R_0$ and varying $L_c$ is treated in the Supplemental Material~\cite{supp}. We can draw a first important counter intuitive conclusion: regardless of the mechanism dictating the inelastic energy loss at the wall of the constriction, the electronic temperature drops when increasing the electrical power dissipated ($P_{N}=V_{\rm b}^2 G_0 N$). Figure~\ref{fig:peakTvsN1} agrees with experimental trend of Fig.~\ref{fig:apdvsG}(b) provided we can link the electron temperature $T_e$ to the number of photons emitted by the contact.

In a bulk metal, non-equilibrium electrons loose their energy mostly during non-radiative collisions with phonons or impurity atoms. Primary photons are emitted as a result of correspondent bremsstrahlung processes. Establishment of thermal equilibrium of photons is a result of complicated kinetics of free-free electron transitions consisting in emission and absorption bremsstrahlung processes as well as Compton effect~\cite{Kompaneets,Zeldovich}. In a simplified diffusion approximation, photon emission can be treated through the radiation transfer equation
\begin{equation}
\frac{dI_{\omega}}{dS}=-\alpha_{\omega}I_{\omega}+\alpha_{\omega}B_{\omega}(T_e)
\end{equation}
where $I_{\omega}$ is the radiation intensity spectrum, $\alpha_{\omega}$ is the absorption coefficient at given frequency, and $B_{\omega}(T_e)$ is the equilibrium radiation intensity given by Plank's law. In a bulk metal, when the optical skin depth $\alpha_{\omega}^{-1}$ is much smaller than the characteristic dimension, Eq.(6) results in the Kirchhoff's law, the emissivity is given by $j_{\omega}=\alpha_{\omega}B_{\omega}(T_e)$. For the interconnection considered here, the region of elevated electron temperature is approximately $r_N\approx N\lambda_F/4<<\alpha_{\omega}^{-1}$. As a result, an equilibrium photon distribution cannot be established within the interconnection region, and the thermal emission of photons is primary guided by a bremsstrahlung process rather than the Kirchhoff's law. The derivation of this bremsstrahlung mechansim is detailed in the Supplemental Material~\cite{supp}.
We finally find the total photon rate 
\begin{equation}
\frac{dN_{ph}(\omega)}{dS d\omega dt} \approx \frac{4e^2\varepsilon_F^2}{3\pi^3 \hbar^3 c^3}\frac{1}{\exp(\hbar\omega/k_B T_e)-1}
\label{eq:17}
\end{equation}

One can easily check that for the temperature domain of interest, i.e. for peak temperatures below $3.5\times 10^3$K (see Fig.~\ref{fig:peakTvsN1}), the integrated bremsstrahlung photon yield is well approximated by the following relation, $dN_{ph}/dSdt \approx 6.13\times 10^{22}\times T^2\exp(-1/T)$~cm$^{-2}$s$^{-1}$, where the normalized temperature $T = k_BT_e/\hbar \omega$.

The results of our calculation of the bremsstrahlung photon yield are shown in Fig.~\ref{fig:peakTvsN1}(c) and (d) as a function of $N$ and for both limiting values of the parameter $h$ governing the heat transfer at the side wall of the system. The data correspond to the calculated peak temperatures shown in Fig.~\ref{fig:peakTvsN1}(a) and (b). The photon yield is calculated by taking into account the APD spectral efficiency (See Supplemental Material~\cite{supp}).  One can find that these dependencies at sufficiently small values of radius $R_0$ qualitatively recover the experimental data shown in Fig.~\ref{fig:apdvsG} notably the exponential decay of the photon counts with $N$. The dependence of the photon yield with fixed $R_0$ and varying $L_c$ is treated in the Supplemental Material~\cite{supp}. We use the model described above to match the experimental dependence of the photon counts versus electrical power delivered in the contact again considering the two extreme heat exchange scenarii at the side wall. The open blue and magenta triangles in Fig.~\ref{fig:apdvsG} are the results of the model considering a short cylinder length $L_c=\lambda_F/4$ and $R_0=2\lambda_F$. We estimate the total radiation area as $S\approx\pi R_0^2+2\pi R_0L_c=4.84\times10^{-14}$~cm$^2$. The overall detection efficiency is experimentally unknown, and we leave this as free parameter $\eta$. To fit the maximum calculated yield with the experimental value for the forth quantum channel, we set $\eta\simeq 0.43$, which means the collection efficiency of the microscope is about 43\%. This is a reasonable value considering the objective's numerical aperture and the presence of a glass substrate concentrating the emitted photons in the high index medium. The measured photon counts are bounded by the two limiting cases of the model indicating the qualitative agreement with the model (see dotted curves in Fig.~\ref{fig:apdvsG}). Hence electron thermalization at the side wall is an important process to consider.

The past research in atomic-size point contacts has provided a firm understanding of the radiofrequency electromagnetic response occurring when the system is driven in the linear regime of low bias voltages (e.g. mV range). Recent reports suggested that electrons transported through the contact with a large kinetic energy ($\sim$eV) may unveil new nonlinear mechanisms of light emission. Our findings showed that photons with energies much higher than the kinetic energy of the electron are emitted during the formation of the atomic contact when the transport becomes ballistic.
By assuming an non-equilibrium electron distribution near the contact, we derive a model relating the electron temperature and the photon yield to the number of quantum channel. Within this model, we assume the presence of a small interconnection region where energy exchange is mainly guided by electrons colliding at the side wall. An anomalous electromagnetic response is emitted in an over-bias spectral domain as a result of a bremsstrahlung process occurring at the boundary of the interconnection region. We derive the quantum-mechanical formula for the rate of this bremsstrahlung photon emission, which in the limit $\hbar\omega \rightarrow 0$ coincides with the classical relation. We find a qualitative agreement between the estimated emission rates and the results of our measurements. Currently, the dynamic leading to the formation of the contact remains too rapid to interrogate the spectrum of the emitted photons. Once we have a reliable strategy to stabilize the number of transport channels, these findings will contribute to the development of integrated electrically-driven optical light sources at atomic length scales.

The work was funded by the European Research
Council Grant Agreement 306772, the CNRS/RFBR collaborative research program number 1493 (RFBR-17-58-150007), the COST Action MP1403 ``Nanoscale Quantum Optics'', the Regional \textit{Excellence} funding scheme (project APEX). A.U. and I.S are thankful to Russian Science Foundation (Grant 17-19-01532) and A.B. for access to the nanofabrication facility ARCEN Carnot financed by the Regional council of Burgundy and la D\'el\'egation R\'egionale \`a la Recherche et \`a la Technologie.

\end{document}

\title{Supplemental Material for\\ ``Effect of quantized conductivity on the anomalous photon emission radiated from atomic-size point contacts''}

\author{Micka\"el Buret}
\affiliation{Laboratoire Interdisciplinaire Carnot de Bourgogne CNRS UMR 6303, Universit\'e de Bourgogne Franche-Comt\'e, 21000 Dijon, France
}%

\author{Igor V. Smetanin}%
\author{Alexander V. Uskov}
\affiliation{Lebedev Physical Institute, Leninsky pr. 53, 119991 Moscow, Russia
}
\author{G\'erard Colas des Francs}
\affiliation{Laboratoire Interdisciplinaire Carnot de Bourgogne CNRS UMR 6303, Universit\'e de Bourgogne Franche-Comt\'e, 21000 Dijon, France
}%
\author{Alexandre Bouhelier}
\email{alexandre.bouhelier@u-bourgogne.fr}
\affiliation{Laboratoire Interdisciplinaire Carnot de Bourgogne CNRS UMR 6303, Universit\'e de Bourgogne Franche-Comt\'e, 21000 Dijon, France
}%

\date{\today}
\maketitle

\section{Experimental methodology}

\begin{figure}
\includegraphics[width=1\columnwidth]{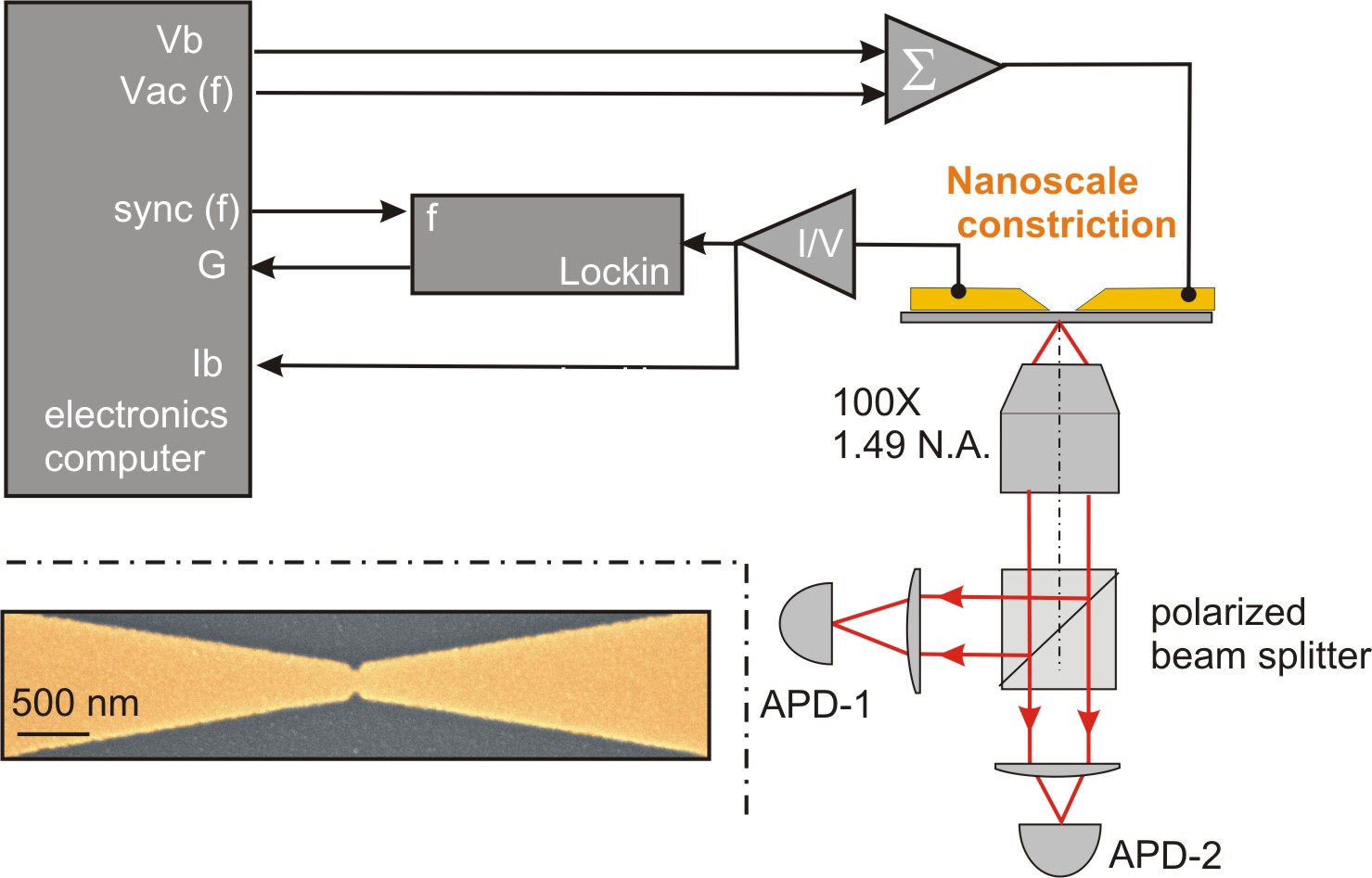}
\caption{\label{fig:setup} Experimental setup used for measuring simultaneously electron transport and light emission during the electromigration of a Au constriction (inset). The conductance of the device is extracted by a lock-in detection and the photons are collected by a high numerical objective and measured with two cross-polarized avalanche photodiodes (APD).}
 \end{figure}

The atomic-size contacts are realized by electromigrating Au constrictions. The constrictions are typically a bow-tie like geometry with a neck width of approximately 150~nm. The constriction and the macroscopic contact electrodes are fabricated by a double step lithography on a glass substrate followed by successive thermal evaporations of a thin layer of Cr and a 50~nm thick layer of Au. The 3~nm thick Cr layer improves the adhesion of gold on the glass. A scanning electron micrograph of a pristine constriction is displayed in the inset of Fig.~\ref{fig:setup}.

The electromigration of the constriction is conducted at ambiant conditions. We apply a variable voltage source $V_{\rm b}$ summed with a 20~mV alternative voltage $V_{\rm ac}$ oscillating at frequency $f=12.1$~KHz. $V_{\rm ac}$ is used to extract the conductance $G$ of the device with a lock-in detection (HF2LI Zurich Instrument). $G=\frac{\partial I_{\rm ac}}{\partial V_{\rm ac}}$, where $I_{\rm ac}$ is the component of the electrical current oscillating at $f$ flowing in the constriction and measured by a current-to-voltage amplifier (I/V DLCPA-200 Femto GmBH). The layout of the experiment is depicted in Fig.~\ref{fig:setup} where signal generation and detection is performed by a scanning electronics (R9 RHK technology). When increasing the voltage $V_{\rm b}$ above the onset of the electromigration, a drop of the conductance signals the reorganisation of the morphology of the conductor. If uncontrolled, this eventually leads to a thermal runaway and a catastrophic rupture of the constriction~\cite{TrouwborstJAPL06}. Instead, if $V_{\rm b}$ is decreased to contain the time evolution of the conductance, the electromigration process is slowed down allowing us to explore the various regimes of electron transport ranging from diffusive to ballistic, and eventually tunnel when the last atomic bond breaks~\cite{Dasgupta:18b}. We then align the constriction to the focus of an inverted optical microscope (Nikon Eclipse) equipped with a high numerical objective (N.A.=1.49) and two photodiode counting modules (APD SPCM-AQR Perkin Elmer).  We use a cross-polarized detection to discriminate photons with an electric field along the main axis of the geometry to those emitted with a transverse polarization state, an expected signature from surface plasmon mediated emission of such biased nanoscale contact~\cite{Uskov_ACS17}. All the experiments are performed at room temperature in a laboratory environment. 

\section{Light emission in the tunnel regime}
Light emitted by tunnel junctions has been a subject of intense research effort since Lambe and McCarthy identified the crucial role of inelastic electron coupling to decaying surface plasmons~\cite{lamb76}.  In the latest advances, tunnel junctions are constituting the active feed of the next generation of electrically-driven optical antennas~\cite{Buret2015,Hecht15,Parzefall15,Liu:18}. In this context, engineering the surface plasmon landscape and the barrier height are expected to boost the notoriously low transduction yield plaguing inelastic energy transfer~\cite{bigourdan16, Uskov16, Parzefall:19}. Continuing on this, a recent proposal suggested that multiple collisions of transported electrons with the boundaries of a plasmonic ballistic constriction may significantly improves the probability to generate an electromagnetic response~\cite{Uskov_ACS17}. 

\begin{figure}
\includegraphics[width=1\columnwidth]{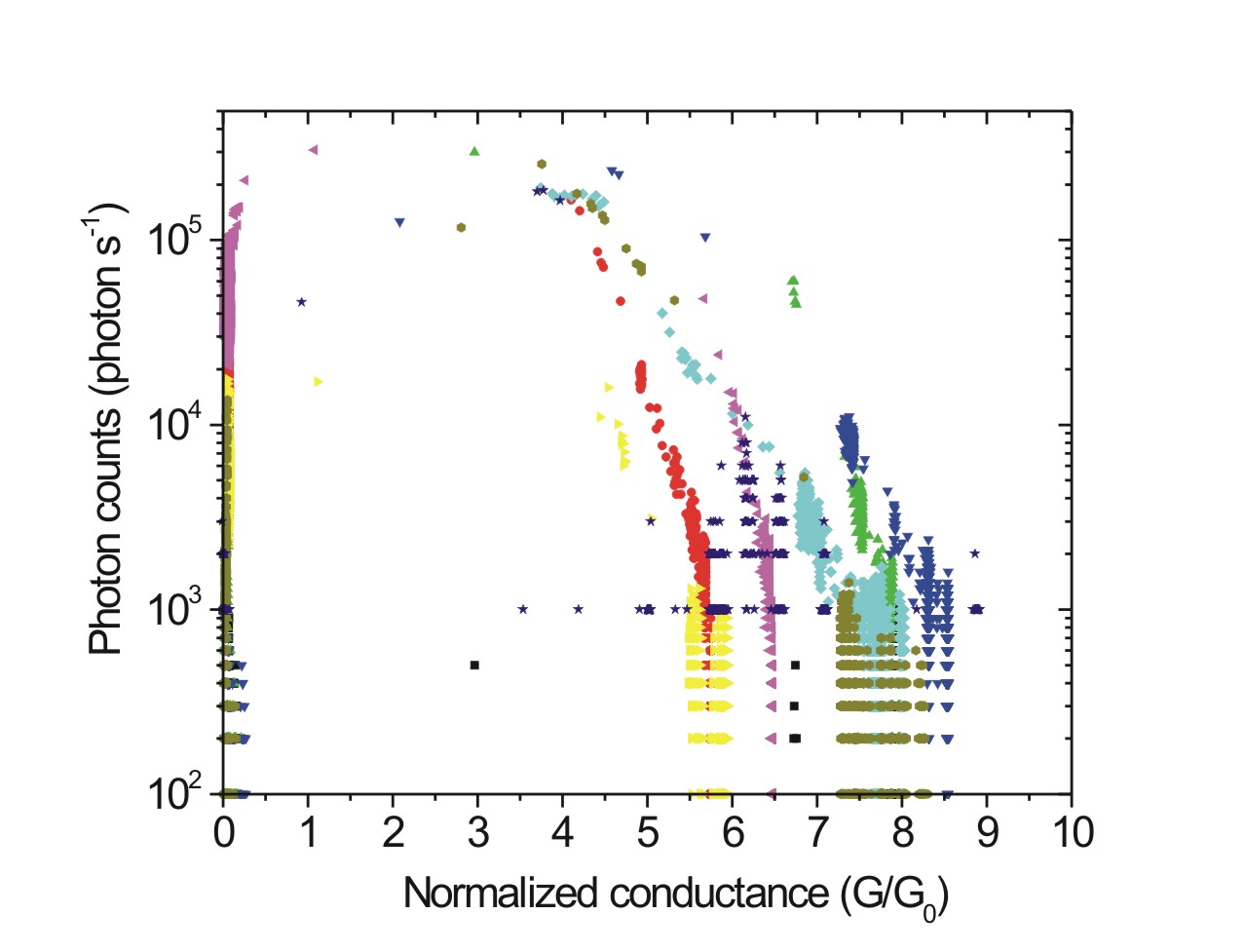}
\caption{\label{fig:apdvsG} (a) Concatenation of results obtained on 9 devices showing the evolution of the photon counts (logarithmic scale) with the normalized conductance. The dark count rate of the APDs is about 300~photon~s$^{-1}$.}
 \end{figure}

As shown in the main text of this article, the radiative pathway triggered by the decay of surface plasmons is not observed experimentally neither in the regime of quantized conduction steps nor when electrons are tunneling. The two cross-polarized signals are at the same count level at all time. Such unpolarized light in the regime of overbias emission confirms our earlier measurement in similar tunnel devices~\cite{Buret2015} and is attributed to the absence of well-defined surface plasmon resonance in this extended geometrical system. In the time traces shown in the main text, light emission is still observe when one of the device operates in the tunneling regime. Considering that $h\nu  \geq eV_{\rm b}$, the emission released in the regime of electron tunneling is either due to the radiative glow of a hot electron distribution~\cite{Buret2015,Dumas16} or due to higher-order electron-plasmon interactions~\cite{Schull09,belzig14,Nitzan15,Berndt17}. The fast dynamics of the last moment of electromigration prevented us to acquire information pertaining to the spectral content of the light, which would have been instrumental for discriminating the physical origin of the light emitted in the regime of electron tunneling. For the second device, the smaller applied bias combined with a lower conductance are inhibiting the emission, if any, to tail in the detected energy window.
The range of conductance values explored here (few $G_0$) before the tunnel barrier forms is similar to the work of Malinowski \textit{et al.} where infrared emission consistent with the black-body radiation of an out-of-equilibrium electron gas was measured in mechanically-controlled break junctions~\cite{Dumas16}. Figure~\ref{fig:apdvsG} is a semi-logarithmic plot displaying the dependence of the total photon counts versus conductance gathered from 9 electromigrated devices. The red circles and the light blue diamonds are data from the two contacts discussed in the main text. The trend is consistent across the tested devices: light emission is detected when the conductance of the contact is entering 8 to 5$G_0$ and dramatically increases up to the breaking point characterized by $G<G_0$. The few data points between $4G_0$ and $G_0$ suggest that the light emission levels off. However, the rapid failure of the contact during the last moment of the electromigration process prevents us to make an affirmative statement. Some devices are also optically active in the tunneling regime as shown by the data points located below $G_0$ in Fig.~\ref{fig:apdvsG}(a).

 \section{Analytical model for electron temperature}
 
At electron temperatures below the Fermi energy $T_e \leq T_F$, the electron thermal conductivity is given by $\kappa_e = C_e v_F^2\tau/3$, $v_F$ is the Fermi velocity and $\tau$ is the characteristic electron transport relaxation time. In a bulk metal, $\tau$ is determined by the electron-phonon momentum relaxtion time and the electron-electron collision rate: $1/\tau=1/\tau_{e-ph}+1/\tau_{e-e}$~\cite{Ashcroft-Mermin:1968,kanavin:98}. When the lattice temperature exceeds the Debye temperature, the electron-phonon collision rate can be estimated as $1/\tau_{e-ph}\sim k_BT_L/\hbar$ so that at the room temperature we find $\tau_{e-ph}\sim$30~fs. Electron-electron collisions dominate at electron temperatures exceeding $T_e\geq T_{*}\sim(E_F T_L/k_B)^{1/2}$, one can find for gold $T_{*}\approx4\times 10^3$~K, where $E_F$ is the Fermi energy and $k_B$ is the Boltzmann constant. In the case of our interest, the actual size of the interconnection region is much smaller than the mean free path of an electron in the bulk material. Hot electrons in the interconnection are quasiballistic: the electron-phonon mean-free path is estimated $\l_{{e-ph}}=v_F\tau_{{e-ph}}\sim 60$~nm for Au~\cite{Chopra63}. The characteristic electron transport relaxation is determined rather by collisions with the walls of the interconnection region, and we can use the following estimate $\tau \sim R_0/v_F$, where $R_0$ is the radius of the interconnection region. As a result, the electron thermal conductivity in the interconnection region is much smaller than that for the bulk material, which provides a large difference between the lattice temperature and the temperature of quasi-ballistic electrons. We will show below that it allows to explain qualitatively our anomalous experimental dependencies of the above-voltage photon yield on the electric power delivered in the contact.

With the above estimate, the electron thermal conductivity coefficient scales linearly with the electron temperature $\kappa_e \equiv \kappa_e (T_e)=b\times T_e$, with the proportionality coefficient $b=\pi^2 v_{F}^2 \tau N_e k_{B}^2/6E_{F}$, $N_e$ is the number of electrons. The electron-lattice coupling constant $g$ can be estimated through the heat capacity of electrons $C_e=C'_eT_e=(\pi^2N_e k_{B}^2/2E_{F})T_e$ and thus also scales linearly with $T_e$, $g\sim C_e/\tau_{e-l}=(C'_e/\tau_{e-l})T_e$. Here $\tau_{e-l}$ is the characteristic time scales for the electron-lattice energy transfer. Keeping in mind these scaling dependences, we can rewrite the electron temperature conduction equation (Eq.~1 in the main text) as $\Delta T_e^2-T_e^2/L_0^2\approx 0$ with the characteristic length $L_0=(v_{F}^2\tau_{e-l}\tau/6)^{1/2}\approx(R_0v_F\tau_{e-l}/6)^{1/2}$.

 \section{Dependence of electron temperature with the geometrical characteristics of the constriction}

\begin{figure}
\includegraphics[width=1\columnwidth]{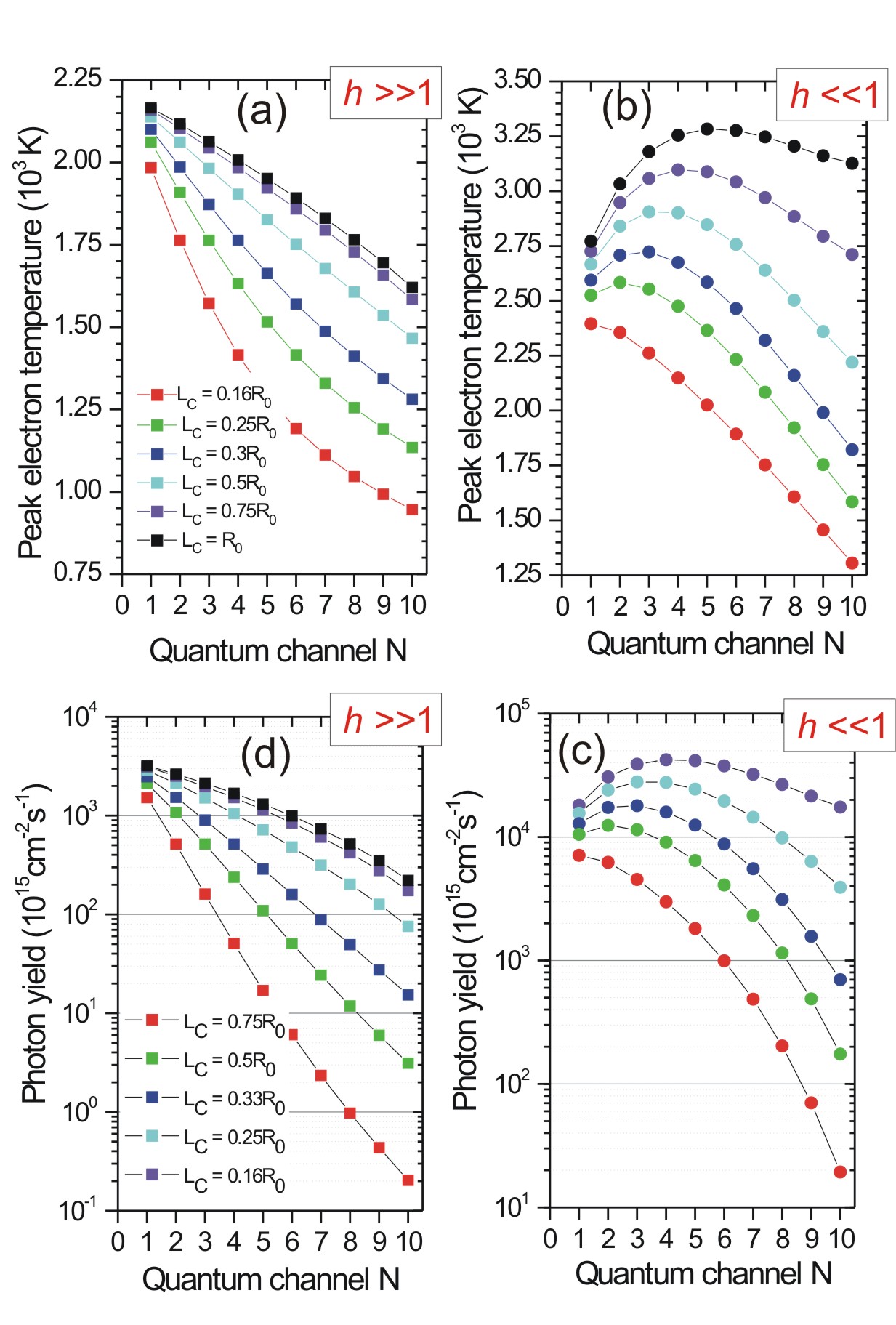}
\caption{\label{fig:peakTvsN1} Peak electron temperature versus the number of open quantum channels $N$ in the contact calculated in (a) and (b) at fixed radius $R_0=2.5 \lambda_F$. (c) and (d) Thermal bremsstrahlung radiation photon rates vs the number of open quantum channels in the contact. Square points represent data at large parameter $h>>1$, while circle points correspond to $h<<1$. Data are plotted for different lengths of the interconnection region: $L_c =R_0/6$ (red), $L_c =R_0/4$ (green), $L_c =R_0/3$ (blue), $L_c =R_0/2$ (cyan), $L_c =3R_0/4$ (violet)  and $L_c =R_0$ (black).}
 \end{figure}

Using the approximate formulations of $T_e$ (Eq.~4 and Eq.~5 in the main text), the dependence of the maximum electron temperature versus the number of quantum transport channels is shown in Fig.~\ref{fig:peakTvsN1}(a) and (b) for both limiting cases of the electron
energy exchange rate at the side wall ($h<<1$ and $h>>1$) and gradually increasing length $L_c$ of the cylinder interconnection region. The radius of the cylinder is fixed at $R_0 =2.5\lambda_F$, which corresponds to a maximum of 10 available quantum channels in the contact. For each channel number $N$, the peak temperature increases with an increase in the cylinder length $L_c$ ranging from $L_c=R_0/6$ to $L_c=R_0$. Obviously, the largest electron temperature is attained when the exchange rate at the side wall is weak ($h<<1$). This dependence is more pronounced for short lengths of the interconnection region when the side wall energy exchange is efficient ($h>>1$). In the other limiting case ($h<<1$), a change in the trend appears with increasing $L_c$  with the occurrence of a maximum shifting to higher $N$. This is the consequence of the first term in brackets in the right hand side of Eq.~5 of the main text, which is linearly growing with $N$. In turns, at sufficiently large $L_c$, the peak temperature will start increasing with $N$ before undergoing a decrease. This is understood from the following argument: 
In the case of small $h$, the only drain of heat is at the distal end of the interconnection, and if the length is sufficiently long, all the transverse oscillations exponentially vanishes according to Eq.~5 except the constant flow (correspondent to the lowest eigen value) which is proportional to the injected power.

 \section{Bremsstrahlung emission emerging from collisions of nonequilibrium hot electrons with the metal boundary}

Digressing from the presence of impurities and from the two-photon Compton emission in electron-electron collisions, we hypothesize that the optical activity detected in our experiment is mainly due to a bremsstrahlung process resulting from hot electrons colliding with the surface potential (at temperatures of our interest one can omit photons emitted in bound-bound transitions in lattice atoms).

To find the photon yield in the electron-wall bremsstrahlung radiation process, we utilize the conventional quantum-mechanical calculation technique, which is analogous to that used in the theory of size-dependent conductivity of thin metal films by Trivedi and Ashcroft~\cite{Ashcroft} as well as in the theory of intersubband transitions in semiconductor quantum wells~\cite{Smeth, Nag}. We consider a metal slab of the thickness $L$, which is considered to be sufficiently large to provide limiting transition to the continuous spectrum of electron momentum. Let the coordinate axis $z$ to be transverse to the slab boundary and $\vec{\rho}$ the coordinate in the boundary plane. The wall of the slab, at $z=0,L$, is modeled by an infinite step-wise potential. Within the jellium model of noninteracting electron system, the wavefunction of an electron inside the slab is
\begin{equation}
\Psi(z,\vec{\rho},t)=\sqrt{\frac{2}{V_e}}\sin[k_z z]\exp(i\vec{k_{\bot}}\vec{\rho})\exp\left(-i\frac{\varepsilon}{\hbar}t\right)
\end{equation}
 which satisfy the boundary conditions $\Psi(z=0)=\Psi(z=L)=0$, $k_z=(\pi /L)j$, $j=1,2,3 \ldots$, is the longitudinal wavenumber, $\hbar \vec{k}_{\bot}$ is the transverse momentum, and $V_e$ is the quantization volume for the electron.  The energy of electrons $\varepsilon= (\hbar^2/2m)(k_z^2+|k_{\bot}|^2)$ is the eigen value of the unperturbed hamiltonian, the perturbation Hamiltonian $H_{int}=-(e/mc)\hat{\vec{A}}\hat{\vec{p}}$, $\hat{\vec{p}}=-i\hbar\nabla $, describes the spontaneous photon emission in a given mode of the wavevector $\vec{k}$, the polarization $\sigma$ and the frequency $\omega =|\vec{k}|c$, with the following vector potential
\begin{equation}
\hat{\vec{A}}=\left( \frac{2\pi \hbar c}{V_{ph} \omega}  \right)^{1/2}\hat{a}_{\vec{k},\sigma}^{+} \vec{e}_{\vec{k},\sigma}^{*}\exp(-i\vec{k}\vec{r})\exp(-i\omega t)
\end{equation}
Here, $\hat{a}_{\vec{k},\sigma}^{+}$ is the correspondent photon creation operator, $\vec{e}_{\vec{k},\sigma}^{*}$ is the unit polarization vector, $V_{ph}$ is the quantization volume for photons. Transitions are induced between the initial state $|i\rangle$ of an elecron with the energy $\varepsilon_i$ and empty photon state and final state $|f\rangle$ of electron with the energy $\varepsilon_f$ and one photon of the above mode. The rate of transition is given by the first order perturbation theory
\begin{equation}
W_{i\rightarrow f}= \frac{2\pi}{\hbar}\left|\langle f\left|H_{int} \right|i \right\rangle |^2 \delta(\varepsilon_f -\varepsilon_i +\hbar\omega)d \rho_f
\label{eq:9}
\end{equation}
Here $d\rho_f$ is the number density of final states, $d\rho_f=(V_e/(2\pi)^3)d^3\vec{k}_f\times (2V_{ph}/(2\pi)^3)d^3\vec{k}$ in the limit of continuous states. The matrix element can be easily calculated as follows,
\begin{eqnarray}
\label{eq:10}
\langle f\left|H_{int} \right|i \rangle & =& i \frac{4e\hbar}{mV_e}\left( \frac{2\pi \hbar c}{V_{ph} \omega}  \right)^{1/2}\times \\ \nonumber
&& \frac{k_{i,z}k_{f,z}}{k^2_{i,z}-k^2_{f,z}}(\vec{e}_{\vec{k},\sigma}^{*},\vec{e}_z)\delta(\vec{k}_{i,\bot}-\vec{k}_{f,\bot})
\end{eqnarray}
Here $\vec{e}_z$ is the unit vector along the $z$ axis. Delta function in the matrix element (Eq.~\ref{eq:10}) demonstrates the conservation of the transverse (parallel to the wall) component of the electron momentum while its $z$-component changes according to the energy conservation law correspondent to the delta-function in the right-hand side of the golden rule (Eq.~\ref{eq:9}). To find the spectrum rate of photon emission by a single electron, we have to make summation over all the electron final states as well as over the polarization and solid angles of photon emission. The sum over the electron final states in the continuous limit $L\rightarrow\infty$ is provided through the following relation
\begin{eqnarray}
&& \int \frac{k^2_{i,z}k^2_{f,z}}{(k^2_{i,z}-k^2_{f,z})^2}\delta^2(\vec{k}_{i,\bot}-\vec{k}_{f,\bot})\times \\ \nonumber
&& \delta(\varepsilon_f -\varepsilon_i +\hbar\omega)d^3\vec{k}_f = (2\pi)^2 S_{\bot}\frac{m^2 v^2_{i,z}v_{f,z}}{4\hbar^3 \omega^2},
\end{eqnarray}
where $S_{\bot}$ is the square of the wall and the velocities $v_{i,z}=\hbar k_{i,z}/m$ and $v_{f,z}=(v^2_{i,z}-2\hbar\omega/m)^{1/2}$ are introduced. The sum over the polarization of bremsstrahlung photons can be accounted through the substitution $2|(\vec{e}_{\vec{k},\sigma}^{*},\vec{e}_z)|^2\rightarrow \sin^2\theta$, and after summation over the photon solid angle $d\Omega=\sin\theta d\theta d\phi$ we arrive in the following relation for the bremsstrahlung emission rate per unit frequency range
\begin{equation}
\frac{dN_i(\omega)}{d\omega dt}=\frac{8e^2 v_{i,z}^2 v_{f,z}}{3\pi V_e \hbar\omega c^3}S_{\bot}
\label{eq:12}
\end{equation}

Let $n_i$ is the number density of electrons with the longitudinal component of velocity $v_{i,z}$, the rate of collisions with the left wall of the metal slab is $\frac{1}{2}n_iS_{\bot}v_{i,z}$ and the same value is the rate of collisions with the right wall of the slab. Consequently, the total emitted power spectrum per one electron ($n_i=1/V_e$) is
\begin{equation}
\frac{dP_{\omega}}{d\omega} = \frac{dN_i(\omega)}{d\omega dt}\frac{\hbar\omega}{v_{i,z}S_{\bot}/V_e}=\frac{8e^2}{3\pi c^3}v_{i,z}v_{f,z}
\end{equation}
In the classical limit $\hbar\omega \rightarrow 0$, i.e., when the energy of scattered electron does not changes significantly, one can replace $4v_{i,z}v_{f,z} \approx (v_{i,z}+v_{f,z})^2=|\vec{v}_i - \vec{v}_f|^2$ and we arrive in the well known classical relation $dP_{\omega}/d\omega=(2e^2/3\pi c^3)|\Delta v|^2$ for the power spectrum emitted by scattered electrons~\cite{Landau2vol, Ghisellini}.

To find the total photon emission rate per unit surface square, we make a summation of Eq.~\ref{eq:12} over all the electron states in the slab assuming the Fermi distribution, $f_F (\varepsilon)=1/( 1+\exp[(\varepsilon-\varepsilon_F)/k_B T_e])$.
 \begin{eqnarray}
  \label{eq:14}
&& \frac{dN_{ph}(\omega)}{dS d\omega dt} = \\ \nonumber
&&=\frac{1}{2S_{\bot}}\int \frac{2d^2\vec{k}_{i,\bot}dk_{i,z}}{(2\pi)^3}V_e \frac{dN_i(\omega)}{d\omega} f_F (\varepsilon_i)(1-f_F (\varepsilon_f)) \\ \nonumber
 &&= \frac{8e^2}{3\pi^3 \hbar^3 c^3}\frac{k_B T_e/\hbar\omega}{\exp(\hbar\omega/k_B T_e)-1}G(\omega,T_e,\varepsilon_F)
 \end{eqnarray}
where $\varepsilon_i=(\hbar^2/2m)(k_{i,z}^2+|\vec{k}_{i,\bot}|^2)$, $\varepsilon_f=\varepsilon_i -\hbar\omega$, the factor $1/2S_{\bot}$ before the integral takes into account doubled scattering surface in the slab. The function $G(\omega,T_e,\varepsilon_F)$ is given by the following integral
\begin{eqnarray}
\label{eq:15}
&& G(\omega,T_e,\varepsilon_F)=\int_{\hbar\omega}^{\infty}du\sqrt{u(u-\hbar\omega)} \times \\
&& \left\{\ln \left[1+\exp \left(\frac{\varepsilon_F-u+\hbar\omega}{k_BT_e}\right)\right]-\ln \left[1+\exp \left(\frac{\varepsilon_F-u}{k_B T_e}\right)\right] \right\} \nonumber
\end{eqnarray}
The total bremsstrahlung photon number spectrum emission rate given by Eqs.~\ref{eq:14} and \ref{eq:15} is a complicated function which we will analyze in details elsewhere. To our particular purpose here we will restrict ourselves by
 the conditions of our experiment where the maximum attainable temperature is well below the energy of collected photons, and we have the following relation between the parameters,
 \begin{equation}
k_B T_e \ll \hbar\omega\ll \varepsilon_F.
\end{equation}
One can easily see that under these conditions the logarithms in braces in Eqs.~\ref{eq:15} vanishes when the argument exceeds $u>\varepsilon_F+\hbar\omega$ and $u>\varepsilon_F$, respectively. At $\hbar\omega<u<\varepsilon_F$, the term in braces is approximately constant and equals $\hbar\omega/k_B T_e>>1$, and at $\varepsilon_F<u<\varepsilon_F+\hbar\omega$ it almost linearly decreases to zero value. As a result, we arrive at the following approximation $G\approx(\varepsilon_F^2\hbar\omega/2k_B T_e)(1+O(\hbar\omega/\varepsilon_F))$ and finally find
\begin{equation}
\frac{dN_{ph}(\omega)}{dS d\omega dt} \approx \frac{4e^2\varepsilon_F^2}{3\pi^3 \hbar^3 c^3}\frac{1}{\exp(\hbar\omega/k_B T_e)-1}
\label{eq:17}
\end{equation}

Compared to the Planck's formula for blackbody radiation,
\begin{equation}
\label{eq:planck}
B_{\omega}(T_e)=\frac{\omega^2}{2\pi c^2}\frac{1}{e^{h\omega/k_BT}-1}
\end{equation}
the rate of bremsstrahlung emission is less by the factor
\begin{equation}
\Upsilon =\frac{8}{3\pi^2}\frac{e^2}{\hbar c} \left(\frac{\varepsilon_F}{\hbar\omega}\right)^2,
\end{equation}
 which under condition of our experiment can be estimated of the order of value as $\Upsilon\sim 0.1$. 

To model the total yield of bremsstrahlung photons detected in our experiment, we integrate the spectrum rate (Eq.\ref{eq:17}) with the transmission function $Q(\omega)$ of the detection optical path which includes the spectral sensitivity of the detector. The APD response restricts the detection efficiency to overbias photon energy tailing in the visible part of the spectrum. We model the spectral response of the APDs by the following function:
\begin{equation}
\label{eq:APD}
Q(\omega)\approx   \left \{
   \begin{array}{r c l}
      0  & = & \omega<\omega_1\\

      0.65 \frac{\omega-\omega_1}{\omega_2-\omega_1} &=&\omega_2>\omega>\omega_1\\
      0.65 & = & \omega_3>\omega>\omega_2
   \end{array}
   \right\}
\end{equation}
Here, $\omega_1$ corresponds to the detection threshold of the detector at a wavelength $\lambda_1=1070$~nm, $\omega_2$ corresponds to the peak of detection efficiency at $\lambda_2=740$~nm, and  $\omega_3$ is taken at $\lambda_3=600$~nm.